# Correlation between image quality metrics of magnetic resonance images and the neural network segmentation accuracy

Rajarajeswari Muthusivarajan[1], Adrian Celaya[1], Joshua P. Yung[1], Satish Viswanath[2], Daniel S. Marcus[3], Caroline Chung[4], David Fuentes[1*]

[1]Department of Imaging Physics, University of Texas MD Anderson Cancer Center, Houston, TX 77030 USA.
[2]Department of Biomedical Engineering, Case Western Reserve University, Cleveland, OH 44106, USA.
[3]Department of Radiology, Washington University School of Medicine, St. Louis, MO 63110 USA.
[4]Department of Radiation Oncology, University of Texas MD Anderson Cancer Center, Houston, TX 77030 USA.
[*]Corresponding Author: DTFuentes@mdanderson.org

**Abstract**
Deep neural networks with multilevel connections process input data in complex ways to learn the information. A network's learning efficiency depends not only on the complex neural network architecture but also on the input training images. Medical image segmentation with deep neural networks for skull stripping or tumor segmentation from magnetic resonance (MR) images enables learning both global and local features of the images. Though medical images are collected in a controlled environment, there may be artifacts or equipment-based variance that cause inherent bias in the input set. In this study, we investigated the correlation between the image quality metrics (IQM) of MR images with the neural network segmentation accuracy. For that we have used the 3D DenseNet architecture and let the network trained on the same input but applying different methodologies to select the training data set based on the IQM values. The difference in the segmentation accuracy between models based on the random training inputs with IQM based training inputs shed light on the role of image quality metrics on segmentation accuracy. By running the image quality metrics to choose the training inputs, further we may tune the learning efficiency of the network and the segmentation accuracy.

**1 Introduction**
The success of deep learning (DL)-based image segmentation tools has facilitated the appreciation for the complex relationship between the quality of input medical imaging data and algorithm performance. Given the black-box nature of DL algorithms, it is unclear whether a direct relationship exists between daily scanner-level quality assurance measures (e.g., signal-to-noise, contrast-to-noise) and algorithm performance. A quality measure that is task specific to DL segmentation would enable the patient specific selection of image data for improved reliability and safe deployment of DL algorithms. Our goal in this study was to evaluate the relationship between common image quality metrics (IQMs) and neural network segmentation accuracy to identify the IQMs most consistently associated with improved neural network segmentation accuracy. We focused our efforts on neuroimaging applications in oncology. Auto-segmentation tools for tumors and normal tissue structures have been shown to improve workflow efficiencies in clinical trials and clinical practice [1]. Reliable skull stripping and tumor segmentation are key steps towards analysis of treatment response.

Skull stripping or brain extraction removes non-brain tissues from an image. Many traditional methods were available prior to the neural network era, including histogram analysis [2], multi-atlas methods [3,4], binarization methods [5], Otsu's method [6], mathematical morphology-based algorithms [7-9], watershed [10], graph cuts [11], etc. All of these have their own merits and drawbacks. In general, they were designed for T1-weighted imaging of healthy brain. Algorithmic performance is expected to vary significantly with varying acquisition protocols and resolutions. DL-based methods such as convolutional neural networks (CNNs) can take either 2D or 3D images as input for skull stripping. In the 2D approach, three 2D CNNs are used to account for the three planes of space to acquire the features from the 2D slices of images [12,13]. Generally, a 2D approach is faster and computationally less expensive than 3D. In 3D medical image segmentation, U-net architecture with fully connected CNNs have been proven more successful than the 2D approach and can recognize both local and global features. The symmetric structure (encoder-decoder) of U-net architecture allows U-nets to extract features from the image during down sampling (encoder, reduces the size of the image) and construct the high-resolution output during up sampling path (decoder). Kleesiek et al. [14] developed a 3D neural network for skull stripping to perform voxel-wise image segmentation from $53^3$ pixels by using seven 3D convolutional layers and one convolutional soft-max output layer. Since it uses only one max-pooling layer, this network is not deep. It can handle one or a combination of different modalities: T1-weighted with and without contrast, T2-weighted, and fluid-attenuated inversion recovery (FLAIR). This network showed high Dice scores and specificity compared to conventional methods on different data sets tested. Hwang et al. [15] performed skull stripping on the Neurofeedback Skull-stripped (NFBS) data set with general 3D-U-net and showed better sensitivity performance compared to Kleesiek's above approach. Isensee et al. [16] deepened the U-net structure by adding residual blocks in the encoder and concatenated with original one. Also, they used a bigger patch size of 128 × 128 × 128 voxels which reconstructs the brain mask precisely. By introducing volumetric dilated convolutions in the U-net architecture, Fedorov et al. reduced the number of parameters and time [17].

The same U-net based fully convolutional neural networks are proven successful on automated tumor segmentation on different types of tumors [18]. Accurate tumor segmentation is important for reliable analysis of quantitative imaging parameters, such as diffusion tensor imaging (DTI), dynamic contrast enhanced MRI, and radiomic feature extraction [19, 20, 21], that may serve as predictive biomarkers for treatment response. For brain tumor segmentation, neural networks should be capable of taking multimodal inputs. The modified U-net takes a stack of multimodal images as one input (input-level fusion) and segments out the multiclass labels [22-25]. But in this method, modality-specific information, or the conflict arising between the different modality images, is left behind. In the lateral-level fusion input method, features from each modality are extracted and fed to the higher-level layer. One such example is dual pathway architectures, where the encoder extracts the features separately then concatenates the features to decode [26-29]. Hu et al. [29] demonstrated a dual pathway DenseNet with fully lateral connections. Its first branch input takes T2-FLAIR and T2 patches, which segments whole tumor, and its second branch takes T1 with T1Gd patches; together these two branches segment each tumor class. Fully lateral connections decrease the number of false-positive and false-negative voxels compared to non-lateral and half-lateral structures. But it is hard to find correlations between the different image segmentation modalities [30]. By using the complementary information of MR images in each modality, segmentation accuracy can be improved [31,32].

The above works with DL algorithms have shown promise in automated segmentation, and in general focused on a modality-specific approach or feature extraction to improve segmentation accuracy. The other important factor is quality of the images used for training, which improves the learning efficiency of the neural network. Training data should include the right representatives for the test data to attain the desired segmentation accuracy. A common theme is that noise and contrast reduction decrease neural network performance [33,34]. Site- or scanner-specific variability of MR images, image resolution, and image artifacts may also affect neural network performance. By running quality control assessments on the input images beforehand, we may fine tune the selection of images for training to reduce the potential for bias. Quality control may also improve the neural network's performance for a wide range of test sets and reduce false positives in the subsequent process, such as brain extraction or tumor prediction. Here, we employed the open-source package MRQy [35] for image quality assessment directly into DL algorithm development. We demonstrated an opportunity to identify the images with similar image quality, which is an important step towards improving the accuracy and reliability of the segmentation task.

## 2 Methods
### 2.1 Network Architecture
We implemented a 3D DenseNet [36] for skull stripping and brain tumor segmentation on brain MRI. Our network is constructed from a composition of convolution and down sampling operations that extract features along a contracting path. An expanding path consists of convolution and up sampling operations with 'long skip' connections to integrate features from the corresponding down sampling operations. Each convolutional operation uses a $3 \times 3 \times 3$ kernel size and is followed by a batch normalization and a rectified linear unit activation function. Four such resolution levels are used. At a given resolution, the feature-maps of all preceding layers are used as inputs, and its own feature-maps are used as inputs into all subsequent layers. After each down sampling block, max pooling layers were used, and the number of feature channels were doubled.

### 2.2 Data sets and Network Training
#### 2.2.1 NFBS Data set
Skull stripping and the MRI Quality Control metrics studies were done on the T1-weighted anatomical MRI scans from the NFBS repository [37]. This data set has 125 MRI scans of patients 21 to 45 years old with clinical and subclinical psychiatric symptoms. For each patient, the repository contains a structural T1-weighted anonymized (de-faced) image, skull-stripped image, and brain mask. Each image has a dimension of $256 \times 256 \times 192$ and a voxel size of $1 \times 1 \times 1$ mm$^3$. The NFBS data set is available for download at http://preprocessed-connectomes-project.org/NFBskullstripped/.

Images were preprocessed by applying the z-score intensity normalization and resizing to $192 \times 192 \times 192$. The learning rate was set as 0.0005, and batch size was set to 8. Input size for the network was $64 \times 64 \times 64$ patches. From each image and brain mask, 16 randomly positioned patches were extracted and passed for each epoch. This ensured that each epoch used a slightly different set of training data in addition to the data shuffling from the k-fold cross validation. Stochastic gradient descent with momentum optimizer with $L_2$ regularization factor was used to optimize the network parameters. Generalized Dice loss function was implemented through the Dice pixel classification layer. Training was performed on an NVIDIA GeForce RTX 8000 with 48 GB RAM.

#### 2.2.2 BraTS data set
For the task of multimodal brain tumor segmentation, we used publicly available data provided by Medical Image Computing and Computer Assisted Interventions society's Multimodal Brain Tumor Image Segmentation Challenge: BraTS 2020 [38]. The database includes multimodal scans (T1-weighted [T1], post-contrast T1-weighted [T1Gd], T2-weighted [T2], and T2-FLAIR) as NIfTI files acquired from multiple institutions. Manual segmentations are available for all images, which contain three labels, namely the GD-enhancing tumor (ET — label 4), peritumoral edema (ED — label 2), and necrotic and non-enhancing tumor core (NCR/NET — label 1). Each modality has a volume of 240 × 240 × 155 which is co-registered to the same anatomical template, interpolated to the same resolution (1 mm$^3$), and skull stripped. The BraTS 2020 training data set is available for download at https://www.med.upenn.edu/cbica/brats2020/registration.html.

For neural network training, all four images were fed together as one 4D input. The learning rate was set to 0.01, and batch size was set to 8. Input size for the network was 64 × 64 × 64 patches. From each image and brain mask, 16 randomly positioned patches were extracted and passed for each epoch. This ensured that each epoch uses a slightly different set of training data in addition to the data shuffling from the k-fold cross validation. Stochastic gradient descent with momentum algorithm with $L_2$ regularization factor was used. Generalized Dice loss function was implemented through the Dice pixel classification layer.

## 2.3 Evaluation metrics

The segmentation accuracy is measured by Dice similarity coefficient (DSC) and Hausdorff distance (HD). Dice similarity coefficient [39] between the predicted mask (P) and its corresponding ground truth mask (R) was calculated using: $D = 2 \cdot |P \cap R| / (|P| + |R|)$, where P is the set of voxels in the predicted mask, R is the set of voxels in the ground truth mask. Hausdorff distance is calculated with simpleitk [40] by computing the distance between the set of non-zero pixels of two images (predicted mask and ground truth mask).

## 2.4 Training based on image quality metrics

Ideally, the trained network should perform well regardless the quality of the input. For clinical implementation the robustness of the algorithm should be such that variable quality of imaging data can be tolerated with reasonable algorithm performance. The preprocessing steps, z-score, and N4ITK bias correction attempt to standardize the image intensity. However, the relative differences in the MRI volumes due to the scanner-specific variation in the image resolution, field-of-view, image contrast, imaging artifacts, inhomogeneity may affect the segmentation accuracy. Recently several automated tools were developed for evaluating variations and relative IQMs [36,41]. We run the data sets (NFBS and BraTS) on MRQy, an open-source quality control tool for extracting IQMs from MR imaging data. It builds on HistoQC Python framework, the digital pathology quality control tool. MRQy can detect the foreground automatically for MR images from anybody region and extract imaging-specific metadata (n = 10). It also generates the following 13 IQM, Variance of the foreground (VAR), Coefficient of variation of the foreground for shadowing and inhomogeneity artifacts (CV), Contrast per pixel (CPP), Peak signal to noise ratio of the foreground (PSNR), Signal to noise ratio with different conditions (SNR1-SNR4), Contrast to noise ratio for shadowing and noise artifacts (CNR), Coefficient of variation of the foreground patch for shading artifacts (CVP), Coefficient of joint variation (CJV), Entropy focus criterion for motion artifacts (EFC), Foreground-background energy ratio for ringing artifacts (FBER)) for an MR image.

To evaluate the role of IQMs for an individual MR image on segmentation accuracy, we retrained the network by taking different data splits based the IQM values. IQM does not have any normative/reference values so, we split the data in the generalized way (Figure 1) by imposing three conditions: 1) ascending order split, i.e., the training set will have the images with smaller IQM values and the test set will have the images with larger IQM values; 2) descending order split, i.e., the training set will have the images with larger IQM values and the test set will have the images with smaller IQM values; and 3) trimmed split, i.e., the middle 80% of the data is taken as training set and the top and bottom 10% of data will be treated as the test set. In all three methods, training and test set ratios are adjusted to be same as 5-fold training.

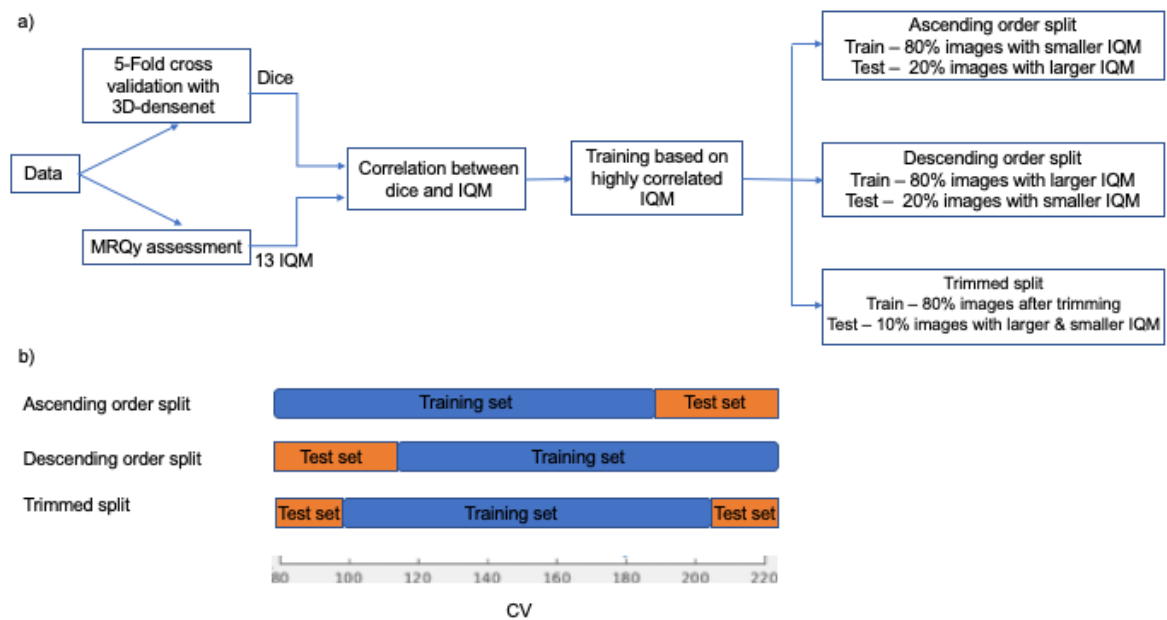

Figure 1. (a) Schematic of data splitting with MRQy image quality metrics (IQM) for training. (b) Data splitting based on the IQM CV.

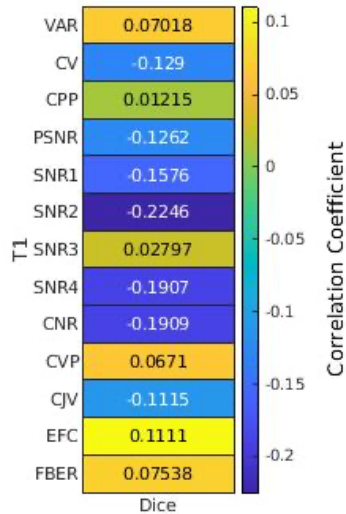

Figure 2. Correlation between dice similarity coefficient and image quality metrics of T1 images.

## 3 Results
### 3.1 NFBS Training and skull stripping based on image quality metrics

Five-fold cross validation was used to train and validate the network. The network trained for 40 epochs, which took around 20 hours. The mean Dice score 0.9709 was comparable with that of similar U-net architecture 0.9903 [15]. The linear dependence of the IQM values with the Dice similarity coefficient (from 5-fold training) was calculated using the Pearson correlation coefficient (Figure 2). For the highly correlated

IQMs (CV, PSNR, SNR1, SNR2, SNR4, CNR, CJV, EFC), network was retrained based on the train-test split as described in the Methods section (2.4). Figure 3 shows the dice scores of retrained networks based on IQM data split. Overall, the Dice scores showed ±0.02 difference compared to the 5-fold Dice score.

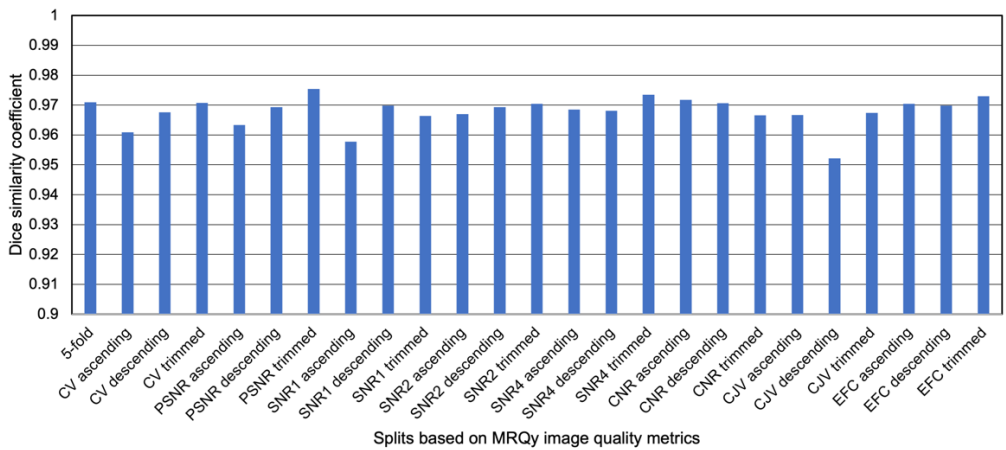

Figure 3. Mean dice similarity coefficient of different train-test data split based on image quality metrics of T1 images.

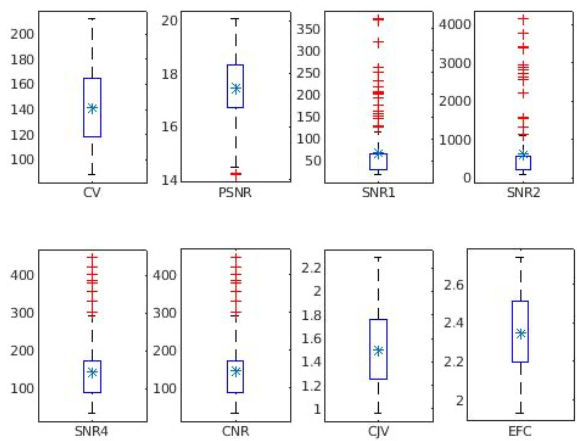

Figure 4. Highly corelated MRQy image quality metrics of T1 images (outliers in red, mean in blue).

Figure 4 shows that the NFBS data had distinct number of outliers for the IQM based on noise measurements such as SNR1, SNR2, SNR4, and CNR. In such cases, training the neural network on the high-quality data alone may not lead to accurate segmentation. The training data set must include representative data for each kind of test data to attain maximum segmentation accuracy. As a proof of concept, SNR1 showed less segmentation accuracy on the ascending order split, where the all the outliers (Larger IQM values of noise measurements represent the better-quality images) were completely get eliminated from the training data set, compared to the other splits (descending and trimmed splits), where they were included. The box plot (Figure 4) of IQMs CV, PSNR, CJV, EFC are showing images having uniform distribution of IQM measure with no significant outliers. For these cases trimmed split shows better segmentation accuracy because the neutral network has seen the right or close representative training images for the test set.

## 3.2 BraTS training and tumor segmentation based on image quality metrics

The network trained for 50 epochs, which took around 80 hours for 5-fold training. For the segmentation evaluation, we generated the whole tumor mask by combining the labels NCR/NET-label 1, ED-label 2, and ET-label 4 and the tumor mask by combining NCR/NET-label 1 and ET-label 4. Dice similarity coefficient for whole tumor, tumor core, and enhancing tumor are shown in Table 1.

Table 1. Mean Dice scores from 5-fold cross validation.

| Method | Mean Dice Score | | |
| --- | --- | --- | --- |
| | Whole tumor | Tumor core | Enhancing tumor |
| 3D DenseNet | 0.7280 | 0.7014 | 0.6726 |
| 3D U-net [38] | 0.9112 | 0.8541 | 0.8058 |

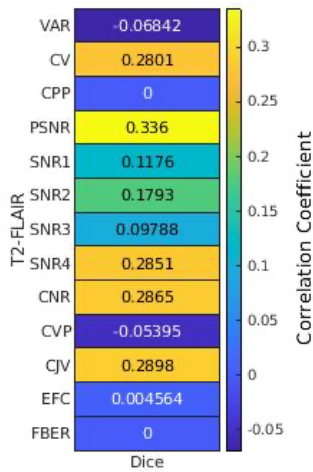

Figure 5. Correlation between the whole tumor dice (5-fold) and the image quality metrics of T2-FLAIR images.

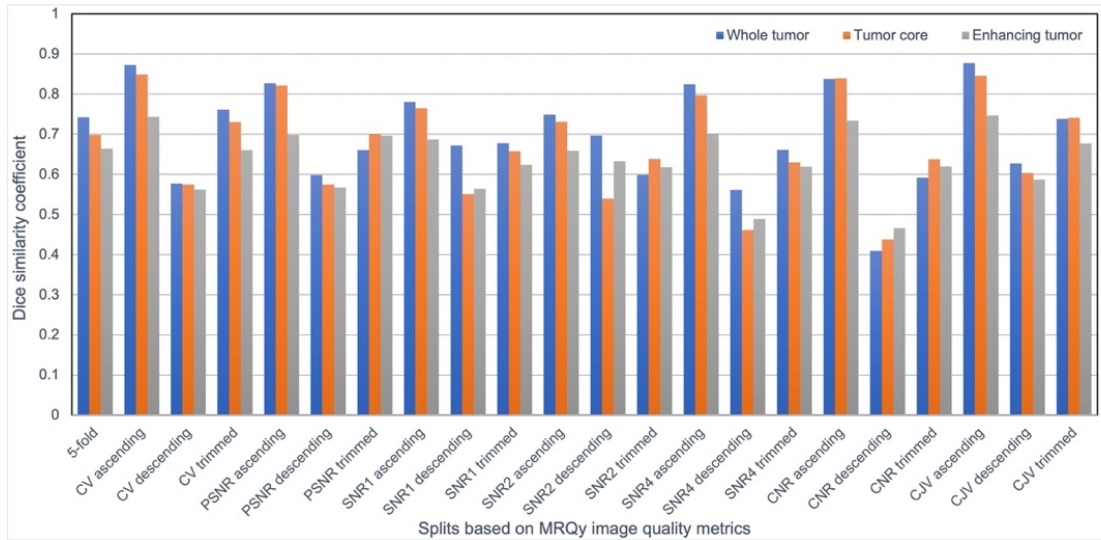

Figure 6. Mean Dice similarity coefficient of different train-test data split based on image quality metrics of T2-FLAIR images.

All four modal images (T1, T1Gd, T2, and T2-FLAIR) of BraTS data were processed through MRQy. Correlations between the MRQy IQM measures and the 5-fold whole tumor dice score were evaluated using the Pearson correlation coefficient. Among the four modalities, T2-FLAIR data shown better correlation with the 5-fold whole tumor dice score (Figure 5). For the IQM-based training 7 highly correlated IQMs (CV, PSNR, SNR1, SNR2, SNR4, CNR, CJV) were selected. Data were split for training and testing based on the T2-FLAIR IQM values (Method section, 2.4). Results of mean dice similarity coefficient and mean 95% Hausdorff distance are compared with 5-fold training (Figure 6 & 7).

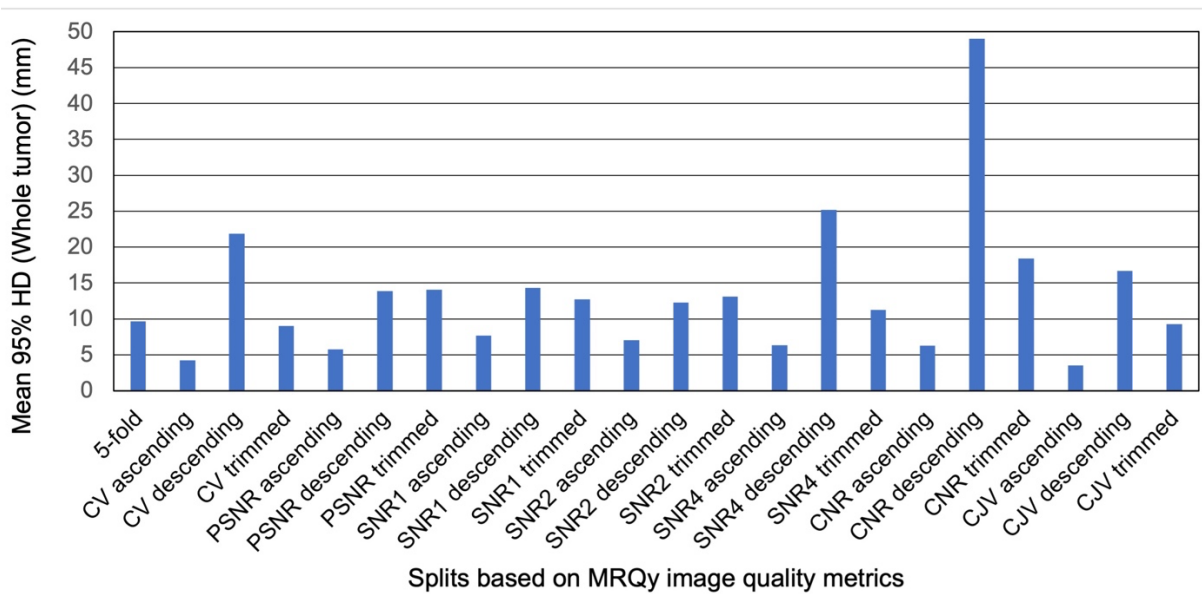

Figure 7. Mean 95% Hausdorff distance (HD) of whole tumor of different train-test data split based on image quality metrics of T2-FLAIR images.

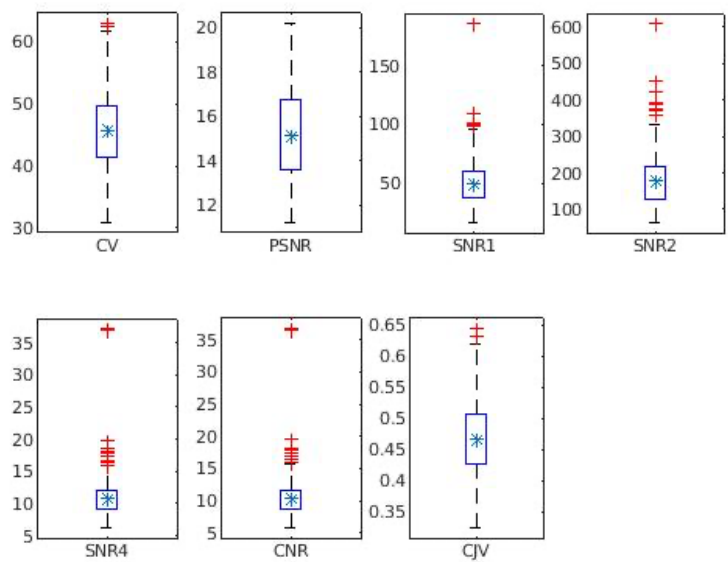

Figure 8. Highly corelated MRQy image quality metrics of T2-FLAIR images (outliers in red, mean in blue).

Overall, there is a significant improvement in the dice scores and Hausdorff distance for ascending order splits than for descending order and trimmed splits. The highest dice scores were obtained for CJV and CV ascending order split trainings compared to 5-fold training and all other IQM splits. In the ascending order split, smaller IQM values were taken for the training set and larger IQM values were in the test set as described in the Methods section (2.4). For both CJV and CV, IQM measure (Figure 8) were distributed uniformly with very few outliers. In MRQy, CJV is calculated between the foreground (F) and background (B) image intensity values (Mean [μ] and standard deviation [σ]): CJV= $\frac{\sigma_F + \sigma_B}{|\mu_F - \mu_B|}$.

The smaller CJV values represent the increased contrast between the tissues [42], and the larger values are related to the presence of heavy head motion and large artifacts [41]. In the CJV descending order split, the training set included the images with larger CJV values, which reduced the overall training image quality compared to the CJV ascending order split. This reduced the dice score, which implies that the high percentage of poor-quality images in the training pool may reduce the segmentation accuracy. Also, the Hausdorff distance was increased considerably for the descending order split, where all the images with high CJV values (poor quality) were in the training pool. But the trimmed split, which included more images with low CJV values (high quality) compared to the descending order split, showed better dice and Hausdorff values. A similar trend was seen for the CV. The IQMs based on noise measurements such as SNR1, SNR2, SNR4, and CNR showed better Dice score and Hausdorff distance for ascending order split than for descending order and trimmed splits.

## 4 Discussion

Our results quantitatively demonstrate that MR image quality plays an important role in a neural network's segmentation accuracy. Our study is similar to a DL study in sensitivity analysis, adversarial robustness as well as intentional and unintentional data-poisoning. Fundamentally, the sensitivity analysis of a neural network is a difficult undertaking. Neural networks are commonly treated as a black box with a high dimensional parameter space. Systematically varying each parameter within a high dimensional neural network to study the effect on the output is expensive. Analysis of a network's sensitivity to inputs such as hyperparameters, regularization parameters, and image quality is obtained by systematically varying the inputs over a meaningful parameter space [43]. Strategies that reduce the network's size to facilitate a practical study the parameter sensitivity may also be used, but it is not clear that results on a smaller network generalize to real-world application. Our approach studies the sensitivity analysis in the relatively small dimension of the image quality. For each IQM considered, the image dimension is effectively reduced to a one-dimensional scalar parameter to characterize and quantify the image quality.

Adversarial perturbations to neural networks have been shown to decrease neural network performance and have received increasing attention within the context of neural network security [44]. Adversarial perturbations may be thought of as noise added to an input image. The adversarial noise may be obtained from the gradient of the loss function likelihood with respect to the imaging input. Intuitively, the loss function gradient is expected to provide the direction of the greatest decrease of the loss function and neural network performance. One might view the deployment of an adversarial attack on a medical imaging neural network system as an unlikely occurrence: An adversarial attacker would need access to the network architecture and optimized weights as well as access to the DICOM servers and PACS systems to perturb the image noise. However, the adversarial noise direction may be relevant in the case when image artifacts or image pre-processing align with the adversarial noise. For example, blurred edges of organ boundaries will tend to align with adversarial noise segmentation problems [45,46]; without clear edges, segmentation fares worse. Future efforts will investigate the relationship between adversarial noise patterns and the IQMs shown to affect image segmentation quality. Intentional data poisoning is also an unlikely scenario. An attacker would need access to the PACS system to modify DICOM images that are being input to a neural network for both training and testing [47]. A large scale of images would need to be manipulated to affect both neural network training and inference. Alternatively, an attacker would need to modify the images within a highly access-controlled public imaging database such as The Cancer Imaging Archive [48].

Further, the neural network would need to be trained on these manipulated images and deployed for inference. In theory, data poisoning efforts could maliciously manipulate images in a way that may be reflected in the IQMs as well. However, within the highly access-controlled setting of medical image analysis, unintentional data poisoning is more realistic. For example, for a neural network trained on a wide variety of diseases, timing, and staging, the imaging may be associated with the disease label itself. Certain diseases would be associated with specific imaging protocols, and site-specific protocol differences are known [35]. Populations associated with these sites could lead a neural network to learn biases associated with image qualities of the specific imaging protocols as well as the scanner manufacturer itself as a surrogate for the underlying patient population. Ambitious patient selection on a large scale would effectively poison the data based on patient subsets. Site-specific acquisition protocol differences are also noted in the IQMs.

Though it is true that the training data set should have a representative of the test set to have a good segmentation accuracy, the overall quality of images in the training pool also a deciding factor for segmentation accuracy. In the NFBS data set, the noise based IQM measurements SNR1, SNR2, SNR4, and CNR had a high number of outliers with larger IQM values/good quality data), mostly pooled in the test set on the ascending order split, which reduced segmentation accuracy (Figure 3). But in the BraTS data, the percentage of outliers is very less, and the IQM measure distribution was more even, which gave the training set combination of good-quality data and representatives of the poor-quality data in the ascending order and trimmed split. This improves segmentation accuracy compared to the descending order split, which included all the poor-quality images (smaller IQM) in the training set (Figure 6). Although, in all cases the dice score was higher for the trimmed split compared to descending order split, which had the middle 80% of data representatives in the training split, ascending order split of inhomogeneity artifacts CV and CJV showed better performance with highest dice score. So, it is evident that the overall quality of the training images and the uniformity in the IQM measure distribution with respect to the test set both have a clear impact on neural network performance.

The k-fold cross validation method with random data splitting ensures that all the observations are used in training and testing. The best scenario is that the segmentation accuracy is similar in all the folds. Though the random train-test data split makes the data set independent and less biased, it is not true for all cases. Because medical images are acquired from various institutions with different site- and scanner-specific variations, imaging artifacts in the individual images reduces neural network performance. Our study shows that by analyzing the IQM of the dataset beforehand, we can fine tune the neural network learning and thus improve the segmentation accuracy with respect to the quality of the individual images.

# 5 Conclusions

In this work, we demonstrated the role of image quality metrics of the MR images in the neural network performance by comparing random 5-fold training and the trainings based on the image quality metrics acquired from MRQy. Our results showed that IQM CV, CJV, PSNR, SNR based trainings shown better segmentation accuracy compared to the random 5-fold training. Segmentation accuracy can be improved by having the uniform data distribution with respect to IQM measures and by having the right representatives for the test set by analyzing the IQM of the dataset beforehand.

**Acknowledgments**

This work was supported through the MD Anderson Strategic Research Initiative Development (STRIDE) Program – Tumor Measurement Initiative.